\documentclass[aps,prl,twocolumn,showpacs,preprintnumbers,amsmath,amssymb]{revtex4}

\usepackage{graphicx}
\usepackage{dcolumn}
\usepackage{bm}

\begin{document}


\title{Geometry and topology of knotted ring polymers in an array of obstacles}

\author{Enzo Orlandini$^{1,2}$}
\author{Attilio L. Stella$^{1,2}$}%
\author{Carlo Vanderzande$^{3,4}$}
\affiliation{%
$^1$Dipartimento di Fisica and CNR-INFM,  Universit\`{a} di Padova, I-35131, Padova, Italy.\\ 
$^2$ Sezione INFN, Universit\`{a} di Padova, I-35131 Padova, Italy.\\
$^3$Departement WNI, Hasselt University, 3590 Diepenbeek, Belgium. \\
$^4$Instituut Theoretische Fysica, Katholieke Universiteit Leuven, 3001 Heverlee, Belgium.
} 


\date{\today}

\begin{abstract}
We study knotted polymers in equilibrium with an array of obstacles which models confinement in a gel or immersion in a melt. We find a crossover in both the geometrical and the topological behavior of the polymer. 
When the polymers' radius of gyration, $R_G$, and that of the region containing the knot, $R_{G,k}$, are small compared to the distance $b$ between the obstacles, the knot is weakly localised and $R_G$ scales as in a good solvent with an amplitude that depends on knot type.   
In an intermediate regime where  $R_G > b > R_{G,k}$, the geometry of the polymer becomes branched. 
When $R_{G,k}$   exceeds $b$, the knot delocalises and becomes also branched. In this regime, $R_G$ is independent of knot type. We discuss the implications of this behavior for gel electrophoresis experiments on knotted DNA in weak fields.\end{abstract}

\maketitle
Circular DNA, as found in bacteria and phages, is often knotted \cite{DNAT05}. Such knots are produced by the action of enzymes known as topoisomerases or by random cyclization inside the head of a phage \cite{Arsuaga}. 
Investigation of the types of knots produced can give important insights in the working of these processes. The knot type of a single DNA can be determined from electron or atomic force microscopy. 
When large numbers of knotted rings have to be analysed one uses gel electrophoresis \cite{Viovy}. 
One expects that for a fixed number of basepairs (bp), a DNA with a more complex knot (as measured for example by the minimal crossing number $n_c$ \cite{complex}) will have a smaller radius and hence, when the applied electric field is weak, a higher mobility. This expectation is verified experimentally \cite{Stasiak} but despite extensive work \cite{Weber06} the precise relation between mobility and knot complexity is not fully understood. For example, the dependence on the length of the DNA, as measured by the number of basepairs, has not been clarified yet. 

The behavior of melts of polymers of various architecture (linear, branched, ring) is also of high interest \cite{Muller96,McLeish08}. In \cite{Obukhov94}, the geometrical and dynamical properties of an {\it unknotted} ring polymer in an array of obstacles were studied. The array was used as a model for a gel but the results were also considered to be relevant for melts of ring polymers. Indeed, in a recent experiment \cite{Kapnistos08}, the relaxation spectrum predicted in \cite{Obukhov94} was observed  in a melt of ring polystyrenes. However, since these polymers were synthesized near theta conditions, they can contain knots \cite{Orlandini07}. This is thus another situation in which topological and geometrical constraints may simultaneously affect the behavior of a polymer. 

Finally, in the crowded environment of a cell \cite{Rivas04}, the average size of a macromolecule is often larger than the average distance between macromolecules, leading, e.g., to important effects on kinetics and geometry of proteins \cite{Homouz08}. Since  some proteins are also knotted \cite{KnotProt} it is of interest to understand how their topological properties are modified in crowded {\it in vivo} situations.

Here we study the behavior of a knotted ring  polymer in a regular, cubic, array of obstacles with lattice constant $b$. We remark that besides being a model of a gel or melt, such regular arrays can also be realised experimentally using hydrogels \cite{Liu99} and in microfluidics \cite{Volkmuth92,Doyle02}, where they have been used to separate DNA. 

In this Letter, we clarify how the excluded volume constraints exerted by the presence of an array of obstacles affect  the geometrical behaviors of knotted ring polymers in equilibrium. The novel insight concerning knot localization \cite{BEAF05} and topology dependence of the size of the rings will be valuable for the interpretation of gel electrophoresis of knotted DNA in weak fields.

Knotted polymer rings are modelled as $N$ edges self-avoiding polygons \cite{Vanderzande98} on the cubic lattice. Each edge corresponds to one persistence length. In the case of DNA this length equals approximately $50$ nm or $150$ bp. 
For an agarose gel, in which most of the experiments with (knotted) DNA are performed, the size of the holes is of the order of $200$ to $500$ nm \cite{Viovy}. 
In our model, this corresponds to a $b$-value from $4$ to $10$. In the microlitographic arrays of ref. \cite{Volkmuth92}, the effective pore size of $1 \mu$m corresponds to $b \approx 20$. 
Vertices occupied by obstacles, and the edges connecting them, are not available for the polygons. This constraint will decrease the overall entropy of the polygons with respect to the free space case affecting both their overall conformational equilibrium properties and the typical size of the knotted portion.

The equilibrium properties of the polygons are studied by Monte Carlo simulations based on BFACF moves \cite{BFACF}. This algorithm is known to be ergodic \cite{Buks91} within the class of fixed knot type and we believe it to be ergodic also for polygons confined within an array of obstacles as long as $b$ is bigger than $2$ lattice spacings. The BFACF algorithm works in a grand canonical ensemble where configurations are weighted with a step fugacity $K$ \cite{Vanderzande98}. The average number of monomers $\langle N \rangle$ will diverge when the fugacity approaches a critical value $K_c$, which is related to the entropy per monomer $s=-k_B \ln K_c$  \cite{Vanderzande98}. To increase the Monte Carlo efficiency we implement the BFACF algorithm on to a multiple Markov chain sampling scheme \cite{Tesi95} using several values of $K$ below the critical value $K_c$. The values of $K_c$ depend strongly on the confining geometry and, for each value of $b$
an exploratory simulation has to be done in order to estimate $K_c(b)$. In free space, $K_c=0.213538$, and from our simulations we determined $K_c(b=10)=0.2227, K_c(b=7)=0.2318$ and $K_c(b=5)=0.2515$.
As expected $K_c(b)$ increases as $b$ decreases since for smaller values of the lattice constant $b$, obstacles are more dense in space, the confinement of the polymers is stronger and hence the entropy $s$ is lower.

\begin{figure}[here]
\vspace{0.4cm}
\includegraphics[width=8.0cm]{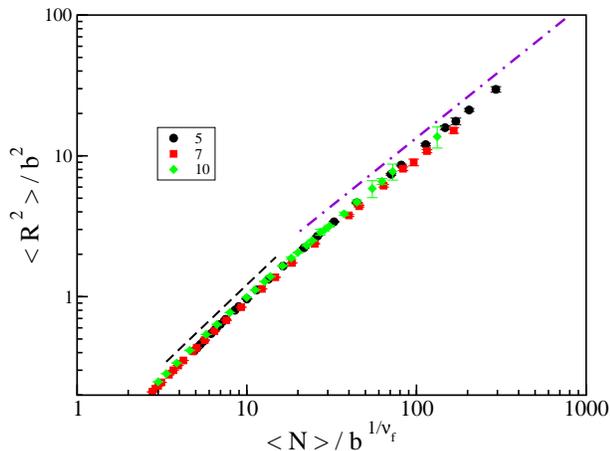}
\caption{\label{fig2} (Color online) 
Log-log plot of the mean squared radius of gyration scaled by $b^2$ as a function of $x=N/b^{1/0.589}$ for three different $b$-values (results for trefoil knot). The dashed line has a slope $1.16$, the dot-dashed one $1.00$.}
\end{figure}

\begin{figure}[here]
\includegraphics[width=8.0cm]{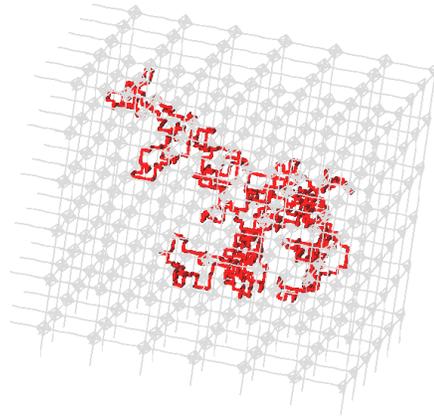}
\caption{\label{fig2/5} (Color online) 
Branched polygon with a composite $3_1 \sharp\ 3_1$ knot in an array with $b=7$.}
\end{figure}

We next determined the average mean squared radius of gyration $\langle R^2\rangle$ as a function of $\langle N\rangle$ for various $b$-values. We expect a power law relation between these two quantities, $\langle R^2\rangle \sim \langle N\rangle^{2\nu}$. In three-dimensional free space, $\nu$ has the value $\nu_f=0.589$. We expect that the polymer begins to 'feel' the presence of the obstacles when its typical size $R_G=\langle R^2\rangle^{1/2}$ becomes of the order of $b$. When $R_G/b$ is small, we should recover the behavior in absence of obstacles. If on the other hand $R_G/b$ is sufficiently large, a new regime dominated by the obstacles should appear. In Fig. \ref{fig2}, we present our data for $\langle R^2\rangle/b^2$ as a function of the scaling variable $x=\langle N\rangle/b^{1/\nu_f}$. 
The results refer to the trefoil knot ($n_c=3$). The data for various $b$-values collapse on a single curve, which can be fitted as a combination of two straight lines, the slopes of which are respectively $2\nu=1.15\pm 0.02$ and $2\nu=0.99\pm 0.02$. 
The former value equals (within the accuracy) the expected value $2\nu_f$. The latter coincides precisely with that for a three dimensional branched polymer (BP) for which $\nu_{BP}=1/2$ \cite{BPd3}.  
That unknotted ring polymers in an array of obstacles should behave as branched polymers was predicted in \cite{Obukhov94}. Our results show that the same is true for knotted polymers and provide the first description  of a polymer that is both branched and knotted (Fig. \ref{fig2/5}). 
The crossover that we find here was never observed before.  
It  occurs around $x=x_c \approx 20$. Fig. \ref{fig2} clearly shows that at the crossover, $R_G \approx b$.

Another important property of knotted polymers is the size $l_k$ of the knotted region. Intuitively it corresponds to the minimal number of monomers that still contain the topological entanglement of the given knot type (for operational definitions of $l_k$, see \cite{BEAF05}).
Research in the last decade has clearly shown that the size of a knot \cite{MHDKK02} is a fluctuating quantity whose statistical properties strongly depends on the physical regime in which the polymer exists \cite{Orlandini09}. In good solvent conditions, it was found that a knot is weakly localized. This means that the average number of monomers in the knot $\langle l_k \rangle$ grows as a power of $N,\ \langle l_k\rangle \sim N^t$, with an exponent $t<1$ \cite{BEAF05}. Numerically, one has $t \simeq 0.75$. In contrast, in the collapsed globular phase, the knot delocalizes ($t=1$) \cite{BEAF05,EAC03}.

\begin{figure}[here]
\includegraphics[width=8.0cm]{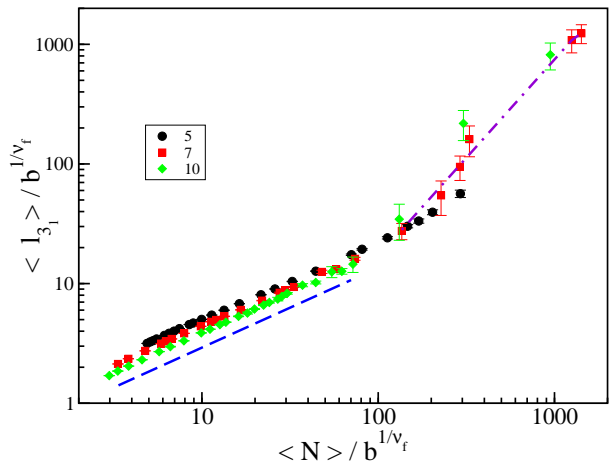}
\caption{\label{fig3} (Color online) 
Log-log plot of the mean knot size scaled by $b^{1/\nu_f}$ as a function of $x$ for three different $b$-values (results for trefoil knot). The blue dashed line has a slope $0.69$, the purple (dot-dashed) one $1.00$.}
\end{figure}

\begin{figure}[here]
\vspace{0.47cm}
\includegraphics[width=8.0cm]{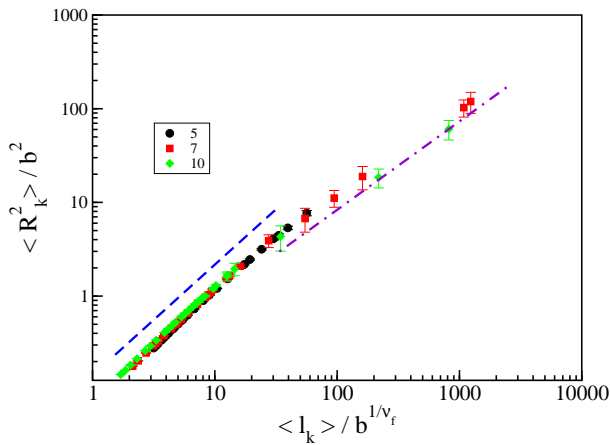}
\caption{\label{fig4} (Color online) 
Log-log plot of the mean squared radius of gyration of the knotted region (scaled by $b^2$) as a function of $\langle l_k\rangle /b^{1/0.589}$ for three different $b$-values (results for trefoil knot). The blue dashed line has a slope $1.18$, the purple (dot-dashed) one $0.98$.}
\end{figure}

For the knotted polymer inside the array of obstacles we determined $l_k$ using the 'cut and join'-algorithm presented in \cite{BEAF05}. In Fig. \ref{fig3}, we report our results for $\langle l_k¾\rangle/b^{1/\nu_f} $ as a function of $x$ for the trefoil for different $b$-values. The results collapse once more on one curve. Moreover, there is again a crossover between two regimes. 
In the first one, the knot is weakly localised with an exponent that we estimate as $t=0.69\pm0.04$, consistent within error bars, with the value in good solvent \cite{BEAF05}. At higher values of $x$, the exponent $t=1.05 \pm 0.15$ indicates a full \emph{delocalization} of the knot.
It is remarkable that this delocalisation crossover occurs at a higher value of $x$ than that of the geometrical crossover, namely near $x=x_d \approx 100$ . This suggest that the delocalisation of the knot only occurs when the average squared radius of gyration of the knotted region, $\langle R^2_k\rangle$ becomes of the order of $b^2$. In Fig. \ref{fig4} we have therefore plotted $\langle R^2_k\rangle/b^2$ versus $\langle l_k \rangle/b^{1/\nu_f}$. These results indeed show that the knotted region behaves in a way completely similar to the whole polymer: once $R_{G,k}=\langle R^2_k\rangle^{1/2}$  becomes equal to the distance between the obstacles, its geometry changes from self-avoiding to branched. Comparing Fig. \ref{fig3} and Fig. \ref{fig4}, we see that the knot indeed delocalises when $R_{G,k}$ is close to  $b$. 

To summarize, there are three regimes for a knotted ring polymer in an array of obstacles. In the first regime the knot is weakly localised and geometrically both the knot and the whole polymer behave as in good solvent with $\nu=\nu_f$. The polymer is essentially confined in one single cell of the array of obstacles. In a second regime, for $20 < x < 100$, the polymer spreads over several cells but the knot is still confined to a single one. As a consequence the geometry of the whole polymer becomes branched, whereas the knotted part still behaves as in good solvent. In these two regimes, the knot is localised. Finally, for $x>100$, the knot delocalises and assumes a branched shape just as the whole polymer.

We next investigated the dependence of our results on knot complexity by extending our calculations to the prime knots $4_1$ and $6_1$ and to the composite knot $3_1 \sharp\ 3_1$ and, for comparison, to the unknot.  For this study we fixed $b=7$.
Firstly, we found that $K_c$ did not depend on the type of knot considered. This result was already established for knotted polygons in free space and our result shows that it is also true within a gel.

\begin{figure}[here]
\includegraphics[width=8.0cm]{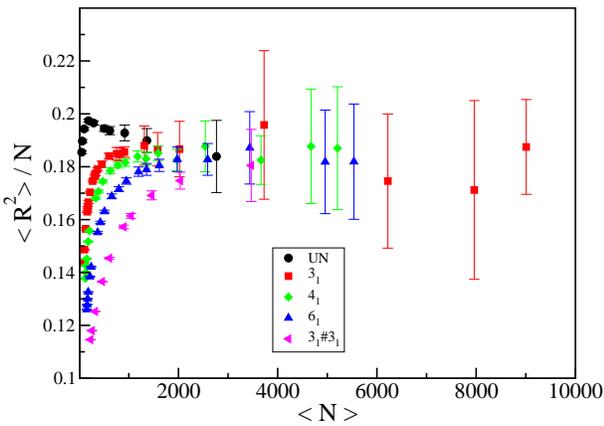}
\caption{\label{fig5} (Color online) 
Average squared radius of gyration per monomer, $\langle R^2\rangle/\langle N \rangle$ versus average number of monomers for different knot types in an obstacle array with $b=7$.}
\end{figure}

\begin{figure}[here]
\vspace{0.35cm}
\includegraphics[width=8.0cm]{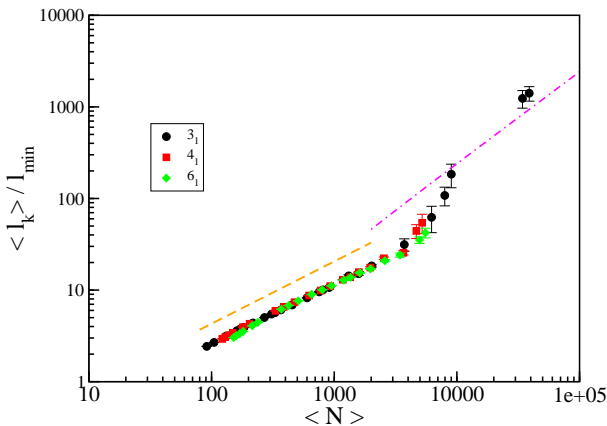}
\caption{\label{fig6} (Color online) 
Average knot size divided by minimum size for the given knot type as a function of $\langle N \rangle$ for three different knot types ($b=7$). The two straight lines have respectively slope 0.69 (orange dashed) and 1.0 (magenta dot-dashed).}
\vspace{-0.25cm}
\end{figure}

In Fig. \ref{fig5} we present our results for $\langle R^2\rangle/\langle N \rangle$ as a function of $\langle N \rangle$ for the various knot types. Plotted in this way, the curves reach a constant in the BP-regime. This figure shows several interesting features. 
Firstly, one observes that for small $\langle N\rangle$, the average squared radius of gyration depends on knot type. The more complex the knot, the smaller is the polymer. 
This is a result similar to that found for ideal knots \cite{Stasiak96}. Secondly, since more complex knots are smaller they feel the presence of the obstacles only for larger $\langle N \rangle$. Hence, $x_c$  depends on knot complexity. Finally, one notices that in the BP-regime, polymer size hardly depends on knot complexity. This result is rather unexpected. Therefore in the BP-regime one can hardly determine the knot topology from geometry. This in contrast to the good solvent regime where the part of the polymer occupied by the knot, as measured by $\langle l_k\rangle$, increases with knot complexity. Evidence for this can be seen in Fig. \ref{fig6} where we plot our results for $\langle l_k\rangle / l_{\min}$ versus $\langle N \rangle$. Here, $l_{\min}$ is the minimal number of edges necessary to embed a given knot type in the cubic lattice.
For the trefoil, the $4_1$ and $6_1$-knot, $l_{\min}$ equals respectively $24, 30$ and $60$ \cite{Diao93}. The data shown in Fig. \ref{fig6} show that more complex knots occupy a larger fraction of the polymer. 
Remarkably, there seems to be an exact proportionality with the minimal length.


Our results can have implications for gel electrophoresis experiments. Assuming as a first approximation that the knotted ring polymer is spherical with radius $R_G$, the mobility in an electric field $\mu \sim R_G^{-1}$, should depend on knot type for small $x$-values. In this regime, different knot types can be separated by electrophoresis. However, our results also predict that for $x$ sufficiently large this is no longer possible.  The genome of the P4-phage is an 11.6 kbp DNA. Its size has been estimated from atomic force microscopy images for DNA with complex knots and was found to be about $300$ nm \cite{Valle05}, which is of the same order as that of the holes in an agarose gel. Similar considerations can be made for the knotted 7 kbp DNA of plasmids. 
While $x_c$ is of course not a universal quantity,  our rough estimates do suggest that the behavior found here is relevant for electrophoresis experiments. Microlithographic arrays of self-assembled magnetic beads \cite{Doyle02} could also be used to reveal the crossovers.

To summarize: for knotted polymers in an array of obstacles, we have found three regimes with distinct geometrical and topological properties. These regimes are manifestations of different degrees of localization of the topological entanglement within the polymer and the array of obstacles.


\end{document}